# Control of fine-structure splitting of individual InAs quantum dots by rapid thermal annealing

D.J.P. Ellis,[a)] R.M. Stevenson, R.J. Young, and A.J. Shields
*Toshiba Research Europe Limited, Cambridge Research Laboratory, 260 Science Park, Milton Road, Cambridge, CB4 0WE, United Kingdom*

P. Atkinson, and D.A. Ritchie
*Cavendish Laboratory, Cambridge University, JJ Thomson Avenue, Cambridge, CB3 0HE, United Kingdom*

**Abstract**

Degeneracy of the bright single exciton spin state is a prerequisite for the production of triggered polarization-entangled photon pairs from the biexciton decay of a quantum dot. Normally, however, the exciton spin states are split due to in-plane asymmetries. Here we demonstrate that the exciton splitting of an individual dot can be tuned through zero by thermal annealing. Repeated annealing blueshifts the exciton emission line of the dot, accompanied by a reduction and inversion in polarization splitting. Annealing is also demonstrated to control the detuning between the exciton and biexciton transitions in any selected dot.

The properties of single semiconductor quantum dots (QDs), their quantized energy levels and ease of integration into more complex structures, has attracted great interest in the field of quantum information[1]. One application is the generation of pairs of polarization entangled photons through the radiative decay of the biexciton state[2-4]. The deployment of QDs in such a source is hindered by the spin splitting of the intermediate exciton level, resulting in only polarization-correlated photon pairs[5-7]. The lifting of the degeneracy of the exciton state is attributed to the physical anisotropies of the quantum dot, such as shape and strain, which break the in-plane symmetry of the electron-hole exchange interaction. This results in linear polarization splitting (S) of both the exciton (X) and biexciton ($X_2$) emission. To create an on-demand entangled photon source, S must be less than the homogenous linewidth of the emission, so that photons are distinguishable by polarization only.

Thermal annealing can be used to blueshift the emission of QD ensembles,[8,9] and control of the average exciton level splitting has been demonstrated through time-domain measurements[10,11]. More recent studies have shown a general relationship between emission energy and S for individual dots in annealed ensemble samples,[12] and that annealing may simultaneously increase the homogenous linewidth[13]. In this Letter we demonstrate that controlled annealing may be used to tune the splitting of a given dot post-growth. We present a systematic study of the spectral evolution of a number of individual QDs which have been annealed in a series of short steps. By analysing the polarization-dependent spectra of the $X_2$ or X emission lines, we are able to determine and map the exciton level splitting of the individual dots as a function of anneal time.

All anneal steps were five minutes in length, and were carried out in a rapid thermal annealer at 675 ºC. This allowed a more reproducible anneal to be achieved compared to shorter, higher temperature anneals. The sample consisted of a single low-density (~1 µm$^{-2}$) layer of self-assembled InAs quantum dots placed between a pair of Bragg mirrors, completely encapsulated within a thin layer of silicon nitride prior to annealing. This was necessary to prevent the desorption of arsenic from the GaAs heterostructure. A high melting point metal was then evaporated onto the sample and patterned with ~ 2 µm diameter apertures. Finally, this film was coated with silicon nitride to ensure that the apertures remained open following annealing.

Before annealing, micro-photoluminescence (µ-PL) was recorded from a number of apertures. All measurements were made at ~5 K in a continuous-flow helium cryostat. The sample was excited using a 532 nm continuous wave laser focussed through a microscope objective lens to a spot size of ~1 µm. Emission was collected through the same objective and analysed with a grating spectrometer and a liquid-nitrogen-cooled charge coupled device. The sample was then annealed and returned to the cryostat, and spectra recorded from the same apertures. In this way we were able to study the spectral evolution of individual QDs as a function of anneal time.

Figure 1 shows PL spectra from two example QDs for different total anneal times. Spectra recorded from dot A (Fig. 1(a)) display two dominant lines, attributed to the X and $X_2$ transitions of the dot based upon the laser power dependence of their intensities[13].

Annealing shifts both lines to higher energy as shown on the figure, where the dashed line is a guide to the eye. The $X_2$ emission redshifts less rapidly than the X, implying a change in the $X_2$ binding energy, $E_X$-$E_{X2}$[14]. As the $X_2$ line crosses the X line and moves to lower energy, the binding energy becomes positive, and the $X_2$ state is said to become bound.

Fig 1(b) shows a series of spectra from a dot for which a number of charged-state emission lines are visible in addition to the X and $X_2$ lines. Such a characteristic line structure is often observed from QDs at these wavelengths[13,16]. As annealing proceeds the $X^+$ line, initially the strongest emission feature, becomes considerably less intense relative to the X line, suggesting that annealing may reduce the excess hole population in the sample. As with dot A, the emission lines move to higher energy with successive anneals and the $X_2$ binding energy becomes inverted – a trend observed for nearly all of the dots. It is also noticeable that the energy separation between the X and $X^+$ states is seen to decrease.

Fig 2(a) illustrates the reduction of S with anneal time for Dot A on a finer energy scale. A polarization splitting is clearly visible in the uppermost (unannealed) X and $X_2$ plots. Each doublet consists of a pair of orthogonally linearly polarized lines, split by 56 ± 3 µeV. As annealing proceeds, S decreases monotonically, becoming negative after the fourth anneal. Changes in exciton emission energy and splitting for both dots are summarised in fig 2(b). Since the initial, $X_2$, and final ground state are spin neutral, the polarization splitting in both the $X_2$ and X emission lines is determined by the splitting of the intermediate exciton state, S, as shown schematically in the inset. To minimise systematic errors, S was determined by averaging the splitting measured in the $X_2$ and X emission, found by Lorenzian fits to the emission lines[12]. From repeated measurements in different cool-down cycles we estimate the accuracy of the measured splitting to be <3μeV. This is larger than during a single cycle[5], due to cycle dependent fluctuations in QD properties, likely from strain and/or charging variation.

Figure 3 shows the average and standard deviation of the X emission energy, X splitting and $X_2$ binding energy for all of the QDs studied, as a function of total anneal time. The average dot emission energy blueshifts after each anneal, while the distribution of dot energies narrows (Fig. 3(a)). A similar trend is observed in the range of measured splittings (Fig. 3(b)), accompanied by an inversion of S for most of the QDs after 20 mins. There is a strong correlation between the emission energy and splitting consistent with Ref 12. However, here all of our measurements are performed on a fixed set of QDs. Figure 3(c) shows the average binding energy at each anneal step, which becomes positive to within one standard deviation, showing that most of the dots have become bound, after 20 minutes of annealing.

A simple explanation for the reduction in S is that annealing reduces the confinement potential of the dot and allows the exciton wavefunction to expand in-plane, reducing the long-range exchange interaction, and hence the splitting[17]. However, this does not explain the observed inversion in the splitting. We believe this originates from the competing influence of shape- and piezoelectric-induced exciton asymmetry. InAs QDs preferentially elongate along the [1 -1 0] direction during growth, resulting in a similar elongation of the exciton envelope[5]. The lattice mismatch between GaAs and InAs results in shear strain which generates piezoelectric potentials within the crystal. This tends to elongate the electron wavefunction in the [110] direction and the hole wavefunction in the [1 -1 0] direction[18]. Thus as annealing proceeds and confinement reduces, the shape-induced asymmetry dominates less as the electron and hole wavefunctions expand in-plane into the barriers. The relative size and extent of the electron and hole states could lead to the elongation of the exciton state in the [110] direction.

In conclusion, we have demonstrated that the emission spectrum and exciton-level splitting of individual QDs can be modified by thermal annealing. We expect that similar control will be possible in the future through localized annealing of an individual dot using a focussed laser beam[19]. The permanent modification to the splitting of a chosen dot we present has significant practical advantages over other, reversible schemes such as applied magnetic[20], electric[21], and strain[22] fields, for which device design and operation is substantially more complicated. Therefore, our results represent an important step towards the reliable production of quantum dots with ~zero exciton level splitting for applications in quantum information.

**Acknowledgements**

The authors would like to acknowledge support from the EPSRC and the European Commission under Framework Package 6 Network of Excellence SANDiE. D.J.P.E. would also like to thank EPSRC and Toshiba for funding and K. Cooper for useful discussions.

**Footnotes**

a) Also at: Cavendish Laboratory, Cambridge University, J.J. Thomson Avenue, Cambridge, CB3 0HE, United Kingdom

**Figure captions**

FIG. 1. (a), (b) Photoluminescence from two different quantum dots, taken after consecutive anneal steps of five minutes duration at 675ºC, showing the evolution of the emission spectrum.

FIG. 2. (a) Polarization dependent PL from Dot A for successive 5 minute anneals at 675°C, plotted relative to the H-polarized peak. Top most spectra is as-grown. Exciton emission is shown in the left-hand panel, biexciton on the right. (b) Summary of the exciton level splitting as a function of exciton energy for dots A and B. Inset depicts energy level structure in a typical dot.

FIG. 3. (a) Evolution of average exciton emission energy of the group of dots studied. (b) Average exciton splitting as a function of anneal time. (c) Average biexciton binding energy as a function of anneal time.

FIG. (a) DOT A / (b) DOT B — PL Intensity (arb) vs Photon Energy (meV), showing peaks labeled $X$, $X_2$, $X^+$, $X_2^+$, $X_2^*$ with increasing anneal time.

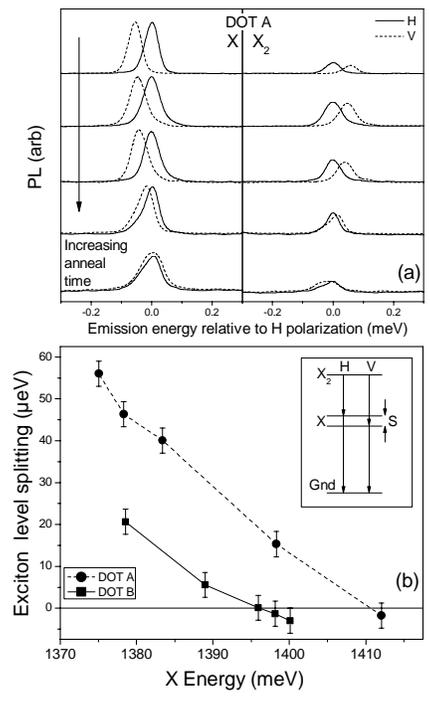

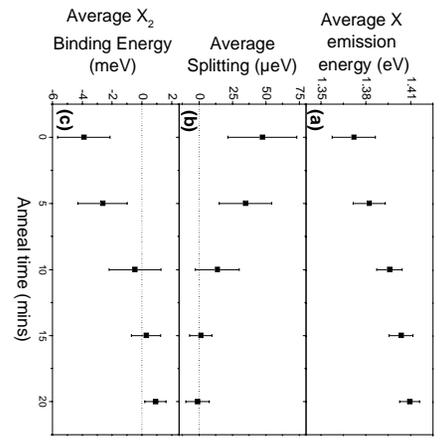